\documentstyle[12pt]{article}
\begin{document}
\rightline{CERN-TH.7428/94}
\rightline{SLAC-PUB-6655}
\vspace{0.7in}
\centerline{\bf ENHANCEMENT OF CP VIOLATION IN ${\bf B^\pm\to
K^\pm_i D^0}$}
\centerline{\bf BY RESONANT EFFECTS}
\vspace{0.5in}
\centerline{\it David Atwood}
\centerline{\it Stanford Linear Accelerator Center, Stanford University}
\centerline{\it Stanford, CA 94309, USA}
\medskip
\centerline{\it Gad Eilam\footnote{
Permanent address:  Department of Physics,
Technion -- Israel Institute of Technology, Haifa 32000, Israel}}
\centerline{\it Department of Physics, Brookhaven National Laboratory}
\centerline{\it Upton, NY 11973, USA}
\medskip
\centerline{\it Michael Gronau$^1$}
\centerline{\it Theoretical Physics Division, CERN}
\centerline{\it CH-1211 Geneva 23, Switzerland}
\medskip
\centerline{and}
\medskip
\centerline{\it Amarjit Soni}
\centerline{\it Department of Physics, Brookhaven National Laboratory}
\centerline{\it Upton, NY 11973, USA}
\vspace{0.5in}
\centerline{\bf ABSTRACT}
\medskip
\noindent
Resonance width effects in charged $B$ decays to neutral $D$ mesons and
excited kaon states $K_i$ around 1400 MeV are shown to lead to large
calculable
final state phases.
$CP$
asymmetries are defined for any charged $B$ decay to three pseudoscalar
mesons involving intermediate overlapping
resonance states. Asymmetries up to about 10$\%$ are found in $B^+\to
K^+_i D^0\to (K\pi)^+ D^0$. Decay distributions can be used to determine
the weak phase $\gamma$ of the Cabibbo-Kobayashi-Maskawa (CKM) matrix.
No separation of the contributions from individual resonances is
required.

\vfill
\leftline{CERN-TH.7428/94}
\leftline{September 1994}
\newpage

In order to observe CP violating effects in a physical process both
a CP violating weak phase and a CP conserving strong phase must
be present \cite{nir}.
In the standard model the weak phase is provided by
phases of the CKM matrix elements \cite{ckm} while the strong phase
must be supplied by the physical system in question.
In the case of neutral $B$ decays the latter phase is
due to the well known oscillation effects.
In order to obtain similar CP violating effects in
$B^\pm$ decay, some other mechanism must be supplied to
provide the strong phase. No evidence for final state phases in
$B$ decays has been found yet \cite{besson}\cite{yamamoto}.
It is possible that these phases are small in two body decays
because of the heavy $B$ meson mass.
In this Letter we shall examine a class of decays of charged
$B$ mesons to hadronic quasi two body final states where the strong
phase is provided by resonant effects in the final state.
In previous papers this approach has been considered in the case of
radiative $B$ decays \cite{atw_soni}.

The processes we wish to study here are of a class suggested some
time ago for a measurement of the weak phase $\gamma$ \cite{gronau},
one of the angles of the CKM matrix unitarity triangle \cite{nir}.
It was shown that this angle may be determined through the rate
measurements of the following processes and their charge-conjugates:
\begin{equation}
B^+\to K^+ \bar D^0, \ \ \ \ B^+\to K^+ D^0, \ \ \ \
B^+\to K^+ D^0_{1,2}.
\label{BtoDK}
\end{equation}
$\bar D^0, D^0$ are the two flavor states, identified
for instance by the lepton charge of their semileptonic decays, and
$D^0_{1,2}=(D^0\pm \bar D^0)/\sqrt{2}$ are the two CP-eigenstates
identified by decay modes such as $K^+ K^-, K_s \pi^0$. In the next
paragraph we explain briefly the method, drawing attention to the role
played by final state phases.

The amplitudes of the first two processes in (1) are governed by CKM
factors $V^*_{cb}V_{us}, V^*_{ub}V_{cs}$. The weak phase difference
between them is $\gamma$. When these amplitudes acquire different final
state phases, the amplitude of the third process, which is their coherent
sum or difference, is expected to show a CP asymmetry with respect to
its charge-conjugate. The angle $\gamma$ is determined from the shape of
the two triangles formed by the magnitudes of the amplitudes of the
three processes in (1) and their charge-conjugates. Although in
principle $\gamma$ can be determined even if the above final state
phases were equal and no CP asymmetry were observed \cite{gronau}, it
would be of great importance to measure a nonzero asymmetry. Also, the
potential accuracy to which $\gamma$ is determined in this way increases
with a growing final state phases difference \cite{stone}.
For instance, a small final state phase difference would inhibit a
useful determination of $\gamma$ if this angle were also small or near
$180^0$. This would correspond to skinny triangles for which the
determination of $\gamma$ becomes quite
challenging. The final state phases are
basically unknown and could be too small for giving rise to
an observable asymmetry in
$B\to K D^0_{1,2}$. In this case it would be useful to find other
related decay channels in which final state phases are enhanced. Here we
will show that when the $K$ meson is replaced by kaonic resonances,
large calculable final state phases are expected to occur. This will not
only improve the prospects of a precise measurement of $\gamma$, but can
potentially also lead to sizable CP asymmetries.

We are thus led to study the process

\begin{equation}
B^+\to K^+_i\bar D^0 , \ \ \ \ B^+\to K^+_i D^0 , \ \ \ \
B^+\to K^+_i D^0_{1,2} ,
\label{BtoDKi}
\end{equation}
in which $K_i$ $(i=a,b,...e)$ are the five lowest lying resonances above
the $K^*(892)$, the properties of which \cite{part_dat} are listed in
Table 1.
Since a strong phase difference is required to obtain an asymmetry,
and since
the dominant effect that we will be focussing on in this work results
from resonance widths, at least two such resonances must decay to a common
final state. The final states in Table 1 are $f=K\pi, K^*\pi, K\rho$.

Consider for illustration the final state $f_1=K\pi$, to which three of
the resonances decay, and let us study the process $B\to K_i D
\to K\pi D$ ($i=c,d,e$).
We define $s=(p_K+p_{\pi})^2=(p_B-p_D)^2$, and denote by $\theta$
the angle between the $B$ momentum and the $K$ momentum in the $K_i$ rest
frame. The amplitudes of the two processes involving $\bar D^0, D^0$
in the final state are
$\bar A^{f_1}_i(s, \theta)$, $A^{f_1}_i(s, \theta)$, respectively.
They are proportional to the
$B^+\to K^+_i \bar D^0$, $B^+\to K^+ D^0$ weak decay
amplitudes: $A(B^+\to K^+_i\bar D^0)=\bar a_i$,
$A(B^+\to K^+_i D^0)=a_i e^{i\gamma}$, respectively. $\bar a_i$
and $a_i$ are assumed to involve small final state phases which will
be neglected.
It should be noted that whereas large final state phases were
measured in two body
$D$ Decays \cite{sheldon}, and sizable phases are required to account
for certain quasi two body $D$ decays \cite{adler}, such phases are
expected to be smaller in decays of the heavier $B$ meson.
We will show later on how this assumption can be tested experimentally
without having to separate the various resonance contributions.
We note
that such phases do not affect the considerations by which $\gamma$ is
determined. Their effect
on the calculated CP asymmetries would
be the appearance of cosines of
these phase differences multiplying the respective interfering
resonant amplitudes.

The two amplitudes, $\bar A^{f_1}_i(s, \theta)$, $A^{f_1}_i(s, \theta)$,
involve a common $s$-dependence,
characterized by the resonance mass and width. We will assume a
(normalized) Breit-Wigner form:
\begin{equation}
\Pi_i(s)={\sqrt{m_i \Gamma_i/\pi}
\over s-m_i^2+i m_i \Gamma_i} .
\label{prop_def}
\end{equation}
The (normalized) $K_i$ decay amplitudes are given by a spin-
characteristic $\theta$ dependence, $\Theta^{f_1}_i(z)$
$(z\equiv\cos\theta)$, multiplied by the square root of the
corresponding decay branching ratios appearing in Table 1,
$\sqrt{B^{f_1}_i}$:
\begin{eqnarray}
\Theta^{f_1}_c&=&\sqrt{{3\over 2}}z, \ \ \ \Theta^{f_1}_d=
{1\over 2}\sqrt{{5\over 2}}(3z^2-1), \ \ \ \Theta^{f_1}_e=\sqrt{{1
\over 2}};
\nonumber\\
B^{f_1}_c&=&0.07, \ \ \ \ B^{f_1}_d=0.50, \ \ \ \ \ \ \ \ \ \ \ \ \ \ \
\ B^{f_1}_e=0.93.
\label{ang_dep}
\end{eqnarray}
The resulting amplitudes, obtained in a narrow width approximation,
are:
\eject
\begin{eqnarray}
\bar A^{f_1}_i(s,z)&\equiv&A(B^+\to K^+_i\bar D^0
\to (K \pi)^+\bar D^0)= \bar a_i\sqrt{B^{f_1}_i}
\Pi_i(s)\Theta^{f_1}_i(z),
\nonumber\\
A^{f_1}_i(s,z)&\equiv&A(B^+\to K^+_i D^0\to (K
\pi)^+ D^0)=e^{i\gamma} a_i\sqrt{B^{f_1}_i}\Pi_i(s) \Theta^{f_1}_i(z).
\end{eqnarray}
The amplitude involving the $CP$-eigenstate $D^0_1$ in the
final state is
\begin{equation}
A^{f_1}_{(1)i}(s,z)={1\over\sqrt{2}}[\bar A^{f_1}_i(s,z)
+ A^{f_1}_i(s,z)] .
\end{equation}
(Of course, a similar expression applies to $D^0_2$).
The total amplitudes for $K\pi$ with invariant mass in the region of
the three overlapping resonances is given by a coherent sum of the
amplitudes through the separate resonances:
\begin{equation}
\bar A^{f_1}=\sum_i \bar A^{f_1}_i, \ \ \
A^{f_1}=\sum_i A^{f_1}_i, \ \ \
A^{f_1}_{(1)}={1\over\sqrt{2}}(\bar A^{f_1}+A^{f_1}).
\end{equation}
We are neglecting a possible nonresonant $K\pi D$
term, the contribution of which in the relatively narrow resonance region
is expected to be suppressed by the ratio of the resonance widths to
$m(B)-[m(D)+m(K)+m(\pi)]$. Note that the first two amplitudes in (7) have
well specified weak phases, their difference being the angle $\gamma$.

In principle, the measured differential decay distributions,
$d^2\bar\Gamma^{f_1}/ds dz$, $d^2\Gamma^{f_1}/ds dz$, $d^2
\Gamma^{f_1}_{(1)}/ds dz$, given by the square magnitudes of the
amplitudes in (7), may be used to extract the three separate resonance
contributions in a partial wave analysis. This would have
 simplified the theoretical study considerably. However, this scenario
may be experimentally difficult due to the limitations of  statistics
and due to the large overlap among the resonances.
We will therefore base our discussion on more realistic considerations,
in which only the combined resonance decay distributions are assumed to
be measurable.

The method of measuring $\gamma$ as described in \cite{gronau} can now
be applied in a differential manner. The third relation of (7) can
be described as a triangle in the complex plane, of which the three
sides represent the three amplitudes. A similar triangle
describes the amplitudes of the charge-conjugated $B^-$ decays, in
which only the sign of the weak phase $\gamma$ has been changed.
The lengths of the sides of the two triangles are given by the square
roots of the differential rates, $\sqrt{d^2\bar\Gamma^{f_1}/dsdz}$,
$\sqrt{d^2\Gamma^{f_1}/dsdz}$, $\sqrt{d^2\Gamma^{f_1}_{(1)}/dsdz}$ and
by the charge-conjugate distributions. This determines the shape of
the two triangles for a given value of $s$ and $\theta$. The angle
between the sides representing $A^{f_1}$ and its charge conjugate is
$2\gamma$. The measured decay distributions for $B^+$
and $B^-$ decays provide a multitude of different pairs of triangles,
all of which share the common angle $2\gamma$. The statistical power
of this method of determining $\gamma$ (leaving out questions of
branching ratios to which we come later) is comparable to the one based
on $B\to KD$, in which a single pair of triangles is determined
from integrated rates. The advantage of excited kaon resonances would
be that large final state phase differences between $\bar A^{f_1}$ and
$A^{f_1}$ occur in the resonance region. As mentioned in the
introduction, this has the effect of improving the precision to which
$\gamma$ can be determined.

Let us proceed to calculate a CP asymmetry. Such an asymmetry is
expected to occur between $d^2 \Gamma^{f_1}_{(1)}/ds dz=
|A^{f_1}_{(1)}|^2$ and
its charge-conjugate, $d^2\bar\Gamma^{f_1}_{(1)}/ds dz$, due to the two
interfering amplitudes, $\bar A^{f_1}$ and $A^{f_1}$, which have
different weak phases and different strong phases in the resonance
region. The method developed below can be applied to calculate CP
asymmetries in any three body hadronic $B$ decay, in which two of
the final particles are decay products of overlapping resonances.

{}From (5)(7) we obtain:
$
$
\begin{equation}
\Delta\equiv
{d^2(\Gamma^{f_1}_{(1)}-\bar\Gamma^{f_1}_{(1)})\over ds dz}=
2\sum_{i,j} (\bar a_i a_j-\bar a_j a_i)(B^{f_1}_i B^{f_1}_j)^{1/2}
\Theta^{f_1}_i\Theta^{f_1}_j {\rm Im}(\Pi_i\Pi^*_j)\sin\gamma.
\label{rate_dif}
\end{equation}
For every pair of resonances this expression has the usual form,
$2A_1A_2\sin(\delta_1-\delta_2)\sin(\phi_1-\phi_2)$, namely a
product of decay amplitudes, a sine of final state phase difference
(here given by ${\rm Im}(\Pi_i\Pi^*_j)$) and a sine of the weak phase
difference ($\gamma$).
Allowing for final state phases in the weak amplitudes would have
introduced cosines of these phase-differences multiplying the
respective products of amplitudes. Another term, involving the sines
of these phases, multiplies the separate Breit-Wigner rates and
integrates to zero in the asymmetries as defined below.

Whereas the $B^+$ $B^-$ partial rate$-$difference
contains only the imaginary
part of interfering resonances, the sum contains both the decay rates
through resonances and the real part of their interference:
\begin{eqnarray}
\Sigma\equiv
{d^2(\Gamma^{f_1}_{(1)}+\bar\Gamma^{f_1}_{(1)})\over ds dz}=
\sum_i (\bar a^2_i +a^2_i +2\bar a_i a_i\cos\gamma)B^{f_1}_i
(\Theta^{f_1}_i)^2 |\Pi_i|^2
\nonumber\\
+2\sum_{i,j}[\bar a_i\bar a_j+ a_i a_j+(\bar a_i a_j + \bar a_j a_i)
\cos\gamma](B^{f_1}_iB^{f_1}_j)^{1/2} \Theta^{f_1}_i\Theta^{f_1}_j
{\rm Re}(\Pi_i\Pi^*_j).
\label{rate_sum}
\end{eqnarray}

The decay distributions (8)(9) have a clearly distinguishable
s-depen\-dence. For simplicity, let us neglect the small mass-differences
among the three relevant resonances and denote the common mass by $m$.
We then find:
\begin{eqnarray}
{\rm Im}(\Pi_i\Pi^*_j)\!\!&\approx&\!\!
{m^2\sqrt{\Gamma_i\Gamma_j}(\Gamma_j-\Gamma_i) \over \pi}
{s-m^2\over [(s-m^2)^2+m^2\Gamma^2_i][(s-m^2)^2+m^2\Gamma^2_j]},
\nonumber\\
{\rm Re}(\Pi_i\Pi^*_j)\!\!&\approx&\!\!
{m\sqrt{\Gamma_i\Gamma_j} \over \pi}
{(s-m^2)^2+m^2\Gamma_i\Gamma_j \over [(s-m^2)^2+m^2\Gamma^2_i][(s-m^2)^2+
m^2\Gamma^2_j]}.
\label{re_im}
\end{eqnarray}
The difference in the  particle-antiparticle differential width
is seen to be an odd function
of $(s-m^2)$, which changes sign at $s=m^2$. On the other hand, the sum
is an even function and contains an interference term which does not
change sign at $s=m^2$. These expressions were obtained when neglecting
final state phases in $\bar a_i, a_i$.
It is easy to show that the sum of the rates
picks up also an odd term if these amplitudes are
allowed to involve final state phases. The absence of such a term in
the measured decay distribution can
be used to test the assumption of negligible non-resonant strong phases.

Since the rate difference changes sign at $s=m^2$ and would vanish
if integrated symmetrically around the resonance mass, it is
useful to define an $s$-integrated asymmetry using the sign function,
$\theta(s-m^2)\equiv +1$ for $s-m^2>0$ and $-1$ for $s-m^2<0$.
In the case under consideration this yields
\begin{eqnarray}
\int ds\theta(s-m^2){\rm Im}(\Pi_i\Pi^*_j)&\approx&
{2\over \pi}{\sqrt{\Gamma_j/\Gamma_i}\over 1+\Gamma_j/\Gamma_i}
{\rm ln}(\Gamma_j/\Gamma_i)
\nonumber\\
&=&\left\{
\begin{array}{lr}
0.25& i=c,\ j=d\cr
0.30&i=e,\ j=d\cr
0.07&i=e,\ j=c\cr
\end{array}
\right. .
\end{eqnarray}
We neglected terms of order $(\Gamma_i/m)^2$ and used the width
parameters of Table 1. The above numbers which describe the integrated
imaginary part of the overlapping resonances could change somewhat with
the resonance parameters but they tend not to be very sensitive to their
precise values. {\it These ``imaginary overlaps'' and similar overlaps
for other pairs of resonances are a key point in any discussion of
resonance effects on final state phases}. In fact, as mentioned above,
they represent the $s$-averaged sines of the final state phase
difference corresponding to two interfering resonances. We see that in
some cases these phases may be large. Also, since their sign is
predicted, the sign of the resulting CP asymmetry will depend only on the
relative magnitudes of certain weak amplitudes.

In the case that the two resonances have different masses,
$m_i\ne m_j$, the corresponding rate difference changes sign at
the $s$-value given by
\begin{equation}
s_0= {m_i^2m_j\Gamma_j-m_j^2 m_i\Gamma_i
\over m_j\Gamma_j-m_i\Gamma_i},
\label{s0_def}
\end{equation}
then a useful $s$-integrated asymmetry will involve the sign
function $\theta(s-s_0)$.

When integrating the particle-antiparticle asymmetry in angular
distribution over all
angles $\theta$ one finds that the partial rate asymmetry vanishes,
unless two intermediate resonances with identical quantum numbers
contribute to the final state \cite{atw_soni}.
 Thus, the usual $z$-integrated asymmetry will project
out the interference of resonances with identical quantum numbers.
To obtain a nonzero asymmetry when the intermediate resonance states
are of different quantum numbers, one may use (similar to the $s$-
integration) a suitable sign function of $z$. Two simple examples of such
functions are:

1. $\theta(z)$, which projects out the interference of resonances
with a unit spin difference.

2. $\theta(|z|-1/2)$, which projects out the interference of spin 0
and spin 2 resonances.

\noindent
In the case under discussion this yields
\begin{eqnarray}
\int dz\Theta^{f_1}_c\Theta^{f_1}_d \theta(z)&=&\sqrt{15}/8=0.48,
\nonumber\\
\int dz\Theta^{f_1}_e\Theta^{f_1}_d \theta(|z|-1/2)&=&3\sqrt{5}/8=0.84,
\nonumber\\
\int dz\Theta^{f_1}_e\Theta^{f_1}_c \theta(z)&=&\sqrt{3}/2\ =0.87.
\end{eqnarray}

We now define two CP asymmetries:
\begin{eqnarray}
{\cal A}_1 &\equiv& {
\int dsdz\theta(s-m^2)\theta(z)d^2(\Gamma^{f_1}_{(1)}-\bar
\Gamma^{f_1}_{(1)})/dsdz \over\Gamma^{f_1}_{(1)}+\bar
\Gamma^{f_1}_{(1)}},
\nonumber\\
{\cal A}_2 &\equiv& {
\int dsdz\theta(s-m^2)\theta(|z|-1/2)d^2(\Gamma^{f_1}_{(1)}-\bar
\Gamma^{f_1}_{(1)})/dsdz \over\Gamma^{f_1}_{(1)}+\bar
\Gamma^{f_1}_{(1)}},
\end{eqnarray}
and calculate them using (8)(9)(11)(13):
\begin{eqnarray}
{\cal A}_1&=& {
[0.045(\bar a_c a_d-\bar a_d a_c)+0.031(\bar a_e a_c-\bar a_c a_e)]
\sin\gamma \over
\sum_i (\bar a^2_i +a^2_i + 2\bar a_i a_i\cos\gamma)B^{f_1}_i},
\nonumber\\
{\cal A}_2&=& {
0.35(\bar a_e a_d-\bar a_d a_e)\sin\gamma \over
\sum_i (\bar a^2_i +a^2_i + 2\bar a_i a_i\cos\gamma)B^{f_1}_i}.
\end{eqnarray}
The first asymmetry is suppressed largely due to the small decay
branching ratio ($7\%$) of the $1^-$ resonances to $K\pi$.
The large numerical coefficient in the second asymmetry demonstrates
the power of this method in enhancing final state phases.
Note that the numerical coefficients depend only on the $K\pi$ decay
branching ratios of the resonances and on the resonance mass-differences
and widths. They do not change by much within the uncertainties in these
experimental quantities. The sign of the asymmetries can be determined
by which of the amplitudes in the parentheses of the numerator
is dominant.

To obtain numerical estimates for these asymmetries one needs to
know $\bar a_i$ and $a_i$, the weak amplitudes into $\bar D^0 K^+_i$
and $D^0 K^+_i$, respectively. In principle, these could be determined
experimentally if the three resonance contributions could be separated.
Some information is also obtained from measuring the total decay rates into
$\bar D^0 K\pi$ and $D^0 K\pi$ final states in the resonance region:

\begin{equation} \bar \Gamma^{f_1}=\sum_i \bar a^2_i B^{f_1}_i,
\ \ \ \ \ \Gamma^{f_1}=\sum_i a^2_i B^{f_1}_i. \end{equation}

Theoretical calculations of the exclusive weak amplitudes are
of course model-dependent.
This is particularly the case for the amplitudes $a_i$, which
are generally expected to be color-suppressed. That is, they involve
$\bar b\to \bar u c \bar s$ quark transitions in which
the $\bar u c$ system is incorporated into a $D^0$ while the $\bar s$
combines with the spectator quark to form the $K^+_i$.
Although color suppression has been observed in $\bar D^0\pi^0,
\bar D^0\rho^0$ final states \cite{besson}, the level of
suppression might vary from process to process and is far from
being understood. Also, to carry out such a calculation,
one would have
to assume factorization and
to rely on quark model wave functions for the
$K_i$ resonances in order to evaluate their weak decay constants.
Two of the kaon resonances under discussion are $q \bar q$
$P$ wave states, for which
weak decay constants have not been much investigated.
Due to all these uncertainties, we will restrict our study to crude
estimates based on an educated guess instead of attempting a model-
dependent calculation. A model-calculation with similar
results will be presented elsewhere \cite{david}.

To estimate $\bar a_c\equiv A(B^+\to K^+_e(1^-)\bar D^0)$, we
use the measured branching ratio for $B^+\to \rho^+ \bar D^0$
of $1.35\%$ \cite{besson}
multiplied by $\sin^2\theta_c=0.22^2$, giving a value
$\bar a^2_c=6.5\times 10^{-4}$. The amplitude of the corresponding $D^0$
mode, $a_c$, is expected to be suppressed both by a color factor,
to be taken as $1/3$ and by the ratio of CKM factors
$|V_{ub}V_{cs}/V_{cb}V_{us}|=0.36$. Correspondingly, we estimate
$a^2_c=9.4\times 10^{-6}$. The other amplitudes of the $2^+$ and $0^+$
$P-$wave states are harder to estimate.
Assuming factorization, the color-
allowed contribution to $K^+_d(2^+)\bar D^0$ is forbidden by angular
momentum conservation.
This state as well as the corresponding one involving a $D^0$
obtain contributions from color-suppressed amplitudes. Thus, we
will assume $\bar a^2_d\approx a^2_d =4\times 10^{-6}$,
a somewhat smaller value than $a^2_c$.
The same value will be taken for $a^2_e$.
The color-allowed amplitude of the $0^+$ state vanishes in the
factorization approximation in a flavor SU(3) symmetry limit in which
the $0^+$ decay constant vanishes. With SU(3) breaking, this decay
constant can be estimated to be given by 0.3 times the $K$ meson decay
constant. A corresponding branching ratio of $\bar a^2_e=2.4\times
10^{-5}$ is then obtained from the measured $BR(B^+\to
\pi^+ \bar D^0)=5.5\times 10^{-3}$ \cite{besson} multiplied by
$\sin^2\theta_c$ and by the SU(3)-breaking factor $0.3^2$.

Thus, we estimate:
\begin{eqnarray}
\bar a^2_c&=&6.5\times 10^{-4}, \ \ \ \ a^2_c=9.4\times 10^{-6},
\nonumber\\
\bar a^2_d&=&4\times 10^{-6}, \ \ \ \ \ \ a^2_d=4\times 10^{-6},
\nonumber\\
\bar a^2_e&=&2.4\times 10^{-5}, \ \ \ \ a^2_e=4\times 10^{-6}.
\end{eqnarray}
With these values we find, for $\gamma=\pi/2$ and for $K\pi$
in the above resonance region:
\begin{equation}
BR(B^+\to (K\pi)^+ D^0_1)=3.8\times 10^{-5}, \ \ \ \ \
{\cal A}_1=1.2\%, \ \ \ {\cal A}_2=2.7\%.
\end{equation}
The two asymmetries are predicted to be positive. In the case of ${\cal
A}_2$ this follows from our estimate that $\bar a_e$ is the dominant of
the four weak amplitudes which are involved in the asymmetry. In the case
of ${\cal A}_1$, in which $\bar a_c$ is the dominant amplitude,
destructive interferences occurs between
the contributions of the two pairs of resonances.
We note that the rather small values
obtained for the asymmetries, in spite of the large final state phases
found in (11), are the result of having two interfering amplitudes which
differ roughly by an order of magnitude.

By choosing the domain
of energy integration more judiciously and/or by using more suitable
weight functions, CP violating asymmetries of about 5 to 10$\%$ are
readily attained \cite{david}. For instance, weighing the asymmetry in
decay distributions by $\Delta/\Sigma$ (defined in (8)(9)) has been shown
to define an optimal asymmetry \cite{atw_soni}. This asymmetry obtains a
value of about 10$\%$ (again, for $\gamma=\pi/2$).

Similar asymmetry calculations were carried out for the two other
resonance decay modes, $f_2=K^*\pi$ and $f_3=K\rho$ \cite{david}.
Asymmetries in the range of $\sim 1$ to $\sim10\%$ are possible.

To summarize, we have shown that in quasi two body $B$ decays
to excited kaon resonances and neutral $D$ mesons (or any other mesons)
large final state phases are generated by the overlapping resonances.
We explained how to use decay distributions to determine the
weak phase $\gamma$. A general definition of (spin-dependent) CP
asymmetries was given, which applies to any quasi two body decay, in
which the dominant final state phases are accounted for in a calculable
manner. Our method does not require resonance separation. Having a good
control over final state phases, the difficult part in CP asymmetry
calculations remains the evaluation of hadronic matrix elements. These
difficulties, though, can get drastically reduced as the branching
ratios for the relevant modes become experimentally known, which is
clearly a much easier task than measurements involving CP violation.
Crude estimates were shown to lead to asymmetries up to
about ten percent. These asymmetries are
the result
of the different orders of magnitudes associated with the two
interfering weak
amplitudes. Clearly, larger asymmetries may be possible in decays
to resonance states in which amplitudes of comparable magnitudes
interfere. Work along these lines is in progress.

\bigskip
\bigskip
\noindent
{\bf Acknowledgements}

We thank A. Ali, D. Cassel and
J. Rosner for fruitful discussions. G. E. wishes to thank the Department
of Physics at Brookhaven National Laboratory and M. G. wishes to thank
the CERN Theory Division for their warm hospitality. This work was
supported in part by the United States $-$ Israel Binational Science
Foundation under Research Grant Agreement 90-00483/3, by the
German-Israeli Foundation of Scientific Research and Development, by an
SSC fellowship and DOE contract DE-AC03-76SF00515 and by DOE contract
DE-AC02-76CH0016.
Partial support (for G. E. and M. G.) was also obtained by the Fund
for Promotion of Research at the Technion.
\eject

\eject
$$
\begin{tabular}{||c|c|c|c|c|c|c|c||}
\hline
Label & Standard Notation & $J^P$ & $m_i$ & $\Gamma_i$ &
$f_1=K\pi$ & $f_2=K^*\pi$ & $f_3=K\rho$\\
\hline
$K_a$ & $K_1(1270)$ & $1^+$ & 1273 &  90  &   -- & 16\% & 42\% \\
$K_b$ & $K_1(1400)$ & $1^+$ & 1402 & 174  &   -- & 94\% &  3\% \\
$K_c$ & $K^*(1410)$ & $1^-$ & 1412 & 227  & $ 7\%$ &$>40$\% &$<7$\% \\
$K_d$ & $K_2(1430)$ & $2^+$ & 1425 &  98  & 50\% & 25\% &  9\% \\
$K_e$ & $K_0(1430)$ & $0^+$ & 1429 & 287  & 93\% &   -- &   -- \\
%$K_f$ & $K  (1460)$ & $0^-$ & 1460 & 250  &   -- & 44\% & 14\% \\
\hline
\end{tabular}
\nonumber
$$
\bigskip\bigskip

\noindent
Table 1:
Properties and branching ratios of $K_i$ resonances. Masses and
widths are given in MeV.
\end{document}